\documentclass[aps,reprint,showpacs,preprintnumbers,amsmath,amssymb,prb,citeautoscript]{revtex4-1} 

\usepackage{bm}
\usepackage{color}
\usepackage{dcolumn}
\usepackage{graphicx}
\graphicspath{{figs/}{./}}
\newcommand\FigureFile[1] {#1.eps}
\DeclareGraphicsExtensions{pdf}

\newcommand\eq[1]                              
{
\begin{align*}
#1
\end{align*}
}

\newcommand\eql[2] 
{
\begin{equation}\label{#1}
\begin{split}
#2
\end{split}
\end{equation}
}

\newcommand\eqsl[1]                            
{
\begin{align}
#1
\end{align}
}

\newcommand\eqssl[2]                      
{
\begin{subequations}\label{#1}
\begin{align}
#2
\end{align}
\end{subequations}
}

\newcommand\Eq[1]      {Eq.~\eqref{#1}}
\newcommand\Eqs[1]     {Eqs.~\eqref{#1}}
\newcommand\Fig[1]     {Fig.~\ref{#1}}
\newcommand\Sec[1]     {Sec.~\ref{#1}}

\newcommand\Ref[1]     {Ref.~\onlinecite{#1}}
\newcommand\Refs[1]    {Refs.~\onlinecite{#1}}

\newcommand\ME[3]      {\langle{{#1}}|{{#2}}|{{#3}}\rangle} 
\newcommand\ket[1]     {|{{#1}}\rangle}

\newcommand\braket[2]  {\langle{{#1}}|{{#2}}\rangle}

\newcommand\PsiGS      {\Psi_0}
\newcommand\PsiT[1][]  {\Psi_{\mathrm{T}#1}^{}}
\newcommand\Dc[1]      {c_{{#1}}^{}}
\newcommand\Cc[1]      {c_{{#1}}^\dagger}

\newcommand\Cphi[1]    {\hat{\phi}_{{#1}}^\dagger}

\newcommand\Half       {\frac{1}{2}}

\newcommand\kvec       {\mathbf{k}}

\newcommand\Hop        {{\hat{H}}}

\newcommand\Hopfc      {\Hop_\Act}
\newcommand\Kop        {{\hat{K}}}

\newcommand\Vop        {{\hat{V}}}

\newcommand\EL         {E_\mathrm{L}^{}}         

\newcommand\Act        {\ensuremath{\mathbb{A}}}
\newcommand\Inact      {\ensuremath{\mathbb{I}}}

\newcommand\Nks        {\ensuremath{N_{\textrm{KS}}}}

\newcommand\compPackage[1] {{\footnotesize{#1}}}

\newcommand\GAMESS     {\compPackage{GAMESS}}
\newcommand\NWCHEM     {\compPackage{NWCHEM}}
\newcommand\ABINIT     {\compPackage{ABINIT}}
\newcommand\ELK        {\compPackage{ELK}}

\newcommand\OPIUM      {\compPackage{OPIUM}}
\newcommand\PWSCF      {\compPackage{PWSCF}}

\newcommand\Ecut       {E_{\mathrm{cut}}}
\newcommand\epsKS      {\Delta\epsilon_{\mathrm{KS}}}  

\newcommand\Order[1]   {\mathcal{O}\left(#1\right)}

\newcommand\Coplus     {Co\textsuperscript{+}}

\definecolor{xmgrace-green4}{rgb}{0.0,0.55,0.0}
\definecolor{Green}{rgb}{0.2,0.96,0.2}
\definecolor{Remarks}{rgb}{1,0.3,0.3}
\definecolor{Extra}{rgb}{0.2,0.2,1}
\definecolor{Blue}{rgb}{0.2,0.3,1}
\definecolor{Black}{rgb}{0,0,0}

\newcommand\COMMENTED[1] {}

\begin{document}

\title{Frozen-orbital and downfolding calculations with auxiliary-field quantum Monte Carlo}

\author{Wirawan Purwanto}
\email{wirawan0@gmail.com}
\affiliation{Department of Physics, College of William and Mary,
Williamsburg, Virginia 23187-8795, USA}

\author{Shiwei Zhang}
\affiliation{Department of Physics, College of William and Mary,
Williamsburg, Virginia 23187-8795, USA}

\author{Henry Krakauer}
\affiliation{Department of Physics, College of William and Mary,
Williamsburg, Virginia 23187-8795, USA}

\date{\today}

\begin{abstract}

We describe the implementation of
the frozen-orbital and downfolding approximations in the
auxiliary-field quantum Monte Carlo (AFQMC) method.
These approaches can provide significant computational
savings compared to fully correlating all the electrons.
While the many-body wave function is never explicit in AFQMC, its random walkers are Slater determinants, whose orbitals
may be expressed in terms of
any one-particle orbital basis.
It is therefore straightforward to partition the full $N$-particle Hilbert space into active and inactive parts
to implement the frozen-orbital method. In the frozen-core approximation, for example, the core electrons can be eliminated in the correlated part of the calculations,
greatly increasing the computational efficiency, especially for heavy atoms.
Scalar relativistic effects are easily included using the Douglas--Kroll--Hess theory.
Using this method, we obtain a way to effectively eliminate the error
due to single-projector, norm-conserving pseudopotentials in AFQMC.
We also illustrate a generalization of the frozen-orbital approach
that downfolds
high-energy basis states to a physically relevant low-energy sector,
which allows a systematic approach to produce realistic model Hamiltonians
to further increase efficiency for extended systems.

\end{abstract}


\maketitle

\section{Introduction}

Many-body methods such as quantum Monte Carlo (QMC) are capable of
providing
the most accurate description of electronic systems
from molecules to extended systems.
Since these methods are significantly more expensive than traditional
mean-field methods such as the density functional theory (DFT),
it is highly desirable to find ways to economize many-body
calculations without sacrificing their predictive power.
Downfolding and partitioning methods have historically been used to develop
effective Hamiltonians, where the inessential degrees of freedom have been eliminated,
so that key aspects  of the correlated systems could be more easily studied.
The Hubbard model exemplifies this approach, albeit being an extreme case since
it removes all materials-specific information.

Theories that partition the Hilbert space into physically important active and inactive subspaces are also well developed for
\emph{ab initio} wave function based, explicitly correlated many-body methods
(see \Refs{Kahn1976} and \onlinecite{Huzinaga1991}, for example).
One of the most widely used example of this partitioning is
 the familiar frozen-core (FC) approximation in quantum chemistry, where many-body
 wave functions are
 expanded as sums of Slater determinants with frozen core orbitals.
This leads to a FC Hamiltonian, which acts only on the subspace spanned by canonical valence and virtual orbitals.
Only the valence electrons are correlated, while the core-valence
interactions appear as one-body operators,
thereby eliminating the core electrons from the calculation.
Closely related valence-only pseudopotential (PP) Hamiltonians,
whose accuracy is based on the validity of the FC approximation,
also invoke this partitioning, but introduce additional approximations.
 Atomic pseudopotentials are
 usually constructed for reference atomic configurations and then used in many target systems.
The accuracy (transferrability) of the PP across many target systems must then
be determined \emph{a posteriori}.
In addition, most PPs used in QMC calculations are of single projector (one per angular momentum channel),
norm-conserving type.
 By contrast, the FC Hamiltonian is obtained
 for each target system, using canonical orbitals from a lower level of theory,
 e.g., the Hartree-Fock (HF) or natural orbitals from a configuration interaction (CI) calculation,
 with no additional algorithmic layers.

It is possible to generalize the frozen orbital approach to other ways
of partitioning the Hilbert space into active and inactive regions.
In molecular and condensed matter physics, for example,
the active region may sometimes be identified spatially,
corresponding to a localized region where strong electron correlation effects
affect a relatively small number of atoms, while the bulk of the system can be treated
with a lower level of theory. This provides opportunities for generating realistic model Hamiltonians
whose many-body treatment will be simpler than the full Hamiltonian but which can retain 
the essential correlation effects quantitatively in a systematic manner.

In this paper we show how downfolding and frozen orbital approaches can be
implemented in the
auxiliary-field quantum Monte Carlo (AFQMC) method
\cite{Zhang2003,AlSaidi2006b,Suewattana2007}
to gain significant computational
savings compared to fully correlating all the electrons.
AFQMC is a many-body method applicable to
condensed matter physics, quantum chemistry, and nuclear physics.
AFQMC stochastically samples the many-body wave function to obtain
observables such as the ground-state energy of a system.
Like other QMC methods, this leads to a modest polynomial scaling
[$\Order{M^3}$ or  $\Order{M^4}$] as the system size is increased,
rather than exponential scaling of CI calculations, or
high-order polynomial scaling of typical quantum chemistry many-body methods.
AFQMC with the phaseless approximation\cite{Zhang2003} has demonstrated
high accuracy in applications to many molecular and extended systems.
\cite{AlSaidi2006b,
Purwanto2009_C2,Purwanto2009_Si,Purwanto2011,%
Ma2012}
AFQMC is based on random walks in the space of Slater determinants,
where each random walker is a full Slater determinant
expressed with respect to a chosen one-particle basis set.
Most AFQMC applications to date have used planewaves for extended systems
and Gaussian-type orbitals (GTO) for atoms and molecules.
In this paper, we show that
the ability of AFQMC to sample explicit Slater determinants and to use \emph{any} one-particle basis
can be exploited to implement various partitioning and downfolding schemes.

The paper is organized as follows:
\Sec{sec:FC} presents the implementation of the frozen orbital approximation
after first reviewing pertinent aspects of the phaseless AFQMC method.
\Sec{sec:FC_atoms} benchmarks the frozen orbital implementation in AFQMC
against exact results in small GTO basis and against experimental results
for large, realistic basis sets.
In \Sec{sec:FC_downfolding}, we demonstrate an application of the downfolding
method to eliminate errors from the use of standard norm-conserving PPs.
\Sec{sec:Summary} summarizes our results and discusses
the prospects of the new frozen orbital capability in AFQMC.

\section{Frozen orbital method in AFQMC}
\label{sec:FC}

AFQMC is an explicitly many-body method for a system of $N$ interacting particles.
The focus here will be on the electronic Hamiltonian for real materials.
Since AFQMC is conveniently formulated in second-quantized form, however,
the methods described in this paper can be directly used to treat any Hamiltonian with one- and two-body interactions.

\subsection{Hamiltonian}

The
Hamiltonian is given by
\eql{eq:H-ae}
{
    \Hop
  &= \Kop + \Vop
  \\&= \sum_{\mu\nu} K_{\mu\nu}^{} \Cc{\mu} \Dc{\nu}
  + \Half\sum_{\mu\nu\lambda\rho}
         V_{\mu\nu\lambda\rho}^{} \Cc{\mu} \Cc{\nu} \Dc{\lambda} \Dc{\rho} \, ,
}
where the lower case Greek indices run over a chosen finite set of
$M$ orthonormal single-particle basis functions;
$\Kop$ and $\Vop$ denote the one- and two-electron interactions, respectively;
and the $\Cc{\mu}$ and $\Dc{\mu}$ are the creation and destruction operators.
This form encompasses all of the orbital-based standard approaches
for interacting electron systems,
from real materials (with all-electron, PP, or FC treatments)
to effective Hamiltonian models (with lattice-based treatments, such as the Hubbard model,
and other downfolded models with reduced degrees of freedom).
Any $N$-electron fermionic state $\Psi$ can be expressed
as a linear combination of Slater determinants
$\ket{\phi} = \Cphi{1} \Cphi{2} \cdots \Cphi{N} \ket{0}$
with their respective weights $a_\phi$,
\eql{eq:Slater-det1}
{
    \ket{\Psi} = \sum_{\phi} a_{\phi} \ket{\phi}
    \,,
}
where $\Cphi{i} = \sum_{\mu} \phi_{\mu i} \Cc{\mu}$.
The number of determinants in this expansion
increases exponentially as a function of $N$ and $M$.
An exact solution of the $N$-electron Hamiltonian is therefore not possible, except for small systems.
AFQMC achieves polynomial scaling by using stochastic methods with importance sampling, as discussed next.

\subsection{AFQMC ground state projection, importance sampling, and phaseless approximation}
\label{sec:FC-AFQMC-review}

This section reviews key elements of AFQMC, which are needed to discuss our implementation of the
frozen orbital approximation.
AFQMC finds the ground state energy $E_0$
using a mixed estimator
and imaginary-time projection from
a trial wave function $\PsiT$
\eql{eq:mixed-Est}
{
 E_0
 =  \frac {\ME{\PsiT}{\Hop}{\PsiGS} }  {\braket{\PsiT}{\PsiGS} }
 =  \lim_{\beta \to \infty} \frac {\ME{\PsiT}{\Hop e^{-\beta \Hop}}{\PsiT} }  {\ME{\PsiT}{e^{-\beta \Hop}}{\PsiT} }
    \,,
}
where $\PsiGS$ is the exact ground state,
and $\PsiT$ is assumed to be non-orthogonal to $\PsiGS$.
The projection in \Eq{eq:mixed-Est} is cast into the form of a branching random walk with Slater determinants $\ket{\phi}$, using
iterations with a small time step $\tau \to 0$,
\eql{eq:gs-proj}
{
    \lim_{\beta \to \infty}  e^{-\beta \Hop} \ket{\PsiT} \approx
    e^{-\tau \Hop}
    e^{-\tau \Hop}
    \cdots
    e^{-\tau \Hop}
    \ket{\PsiT}
    \to
    \ket{\PsiGS}
    \,.
}
The small imaginary time step $\tau$ allows
a Trotter-Suzuki breakup of the exponential operator,
\eql{eq:Trotter-Suzuki}
{
    e^{-\tau \Hop}
    \approx
    e^{-\tau \Kop / 2}
    e^{-\tau \Vop}
    e^{-\tau \Kop / 2}
    + \Order{\tau^3}
    \,.
}
After a Hubbard-Stratonovich transformation of $e^{-\tau \Vop}$,
the projection operator can then be expressed as  a high-dimensional
integral over auxiliary fields $\bm{\sigma}$ \cite{Zhang1997_CPMC,Zhang2003},
\eql{eq:HS1}
{
    e^{-\tau \Hop}
    =
    \int d\bm{\sigma} P(\bm{\sigma})
e^{-\tau {\hat K}/2}
\:e^{\sqrt{\tau} \bm{\sigma} \cdot {\hat{\mathbf v}}}
\:e^{-\tau {\hat K}/2}
    \,,
}
where $P(\bm{\sigma})$ is the normal distribution function,
$\hat{\mathbf v}$
is a one-body operator, and the operator in the integrand, acting on a Slater determinant,
simply yields another determinant,
\eql{eq:step}
{
 e^{-\tau {\hat K}/2}
\:e^{\sqrt{\tau} \bm{\sigma} \cdot {\hat{\mathbf v}}}
\:e^{-\tau {\hat K}/2} \ket{\phi}
\equiv e^{-\tau {\hat h}(\bm{\sigma})} \ket{\phi} \to \ket{\phi'}
\,.
}
Starting with an initial population of walkers (which are usually set equal to $\PsiT$),
the mixed estimator  in \Eq{eq:mixed-Est} is then stochastically sampled.
As the one-body operator ${\hat h}(\bm{\sigma})$ is generally complex, however, the orbitals in $\ket{\phi}$
will become complex as the projection proceeds, and
the statistical fluctuations in the mixed estimator
increase exponentially with projection time.
To control this problem, a phaseless approximation was introduced\cite{Zhang2003}
based on the complex importance function $\langle \Psi_T|\phi\rangle$.
After the importance sampling transformation,
 \Eq{eq:step} becomes
\eql{eq:step-imp}
{
e^{-\tau {\hat K}/2}
\:e^{\sqrt{\tau} \left (  \bm{\sigma} - \bm{\bar{\sigma}}[\phi] \right )\cdot {\hat{\mathbf v}}}
\:e^{-\tau {\hat K}/2}  \ket{\phi} \to \ket{\phi'} \, ,
}
where the ``force bias'' $\bm {\bar{\sigma}}[\phi] $ is given by
\eql{eq:HS-FB}
{
   \bm {\bar{\sigma}}[\phi]
    \equiv
    -\sqrt{\tau}\,
    \frac{\ME{\PsiT}{\hat{\mathbf v}}{\phi}}{\braket{\PsiT}{\phi}}
    \,.
}
The mixed estimator at each time slice becomes a weighted sum over the walkers
\eql{eq:Emixed-walkers}
{
E_0 \approx  \frac {\sum_\phi w_\phi  \EL[\phi]} {\sum_\phi w_\phi} \, ,
}
where
\eql{eq:Elocal-def}
{
    \EL[\phi] \equiv {\ME{\PsiT}{\Hop}{\phi}}/{\braket{\PsiT}{\phi}}
}
is the ``local energy''
of each walker and its weight $w_\phi$ is accumulated over the random walk, as described
in more detail in
\Refs{Zhang2003,AlSaidi2006b,Purwanto2004,Suewattana2007}.

\subsection{AFQMC implementation of the frozen orbital method}

It is often useful to partition the $N$-electron Hilbert space
into active (\Act) and inactive (\Inact) parts, reflecting
physical considerations
based on energetic, spatial, or other factors.
Since AFQMC is an orbital-based method,
this partitioning is facilitated by the
freedom to use {\em any} orthonormal basis in \Eqs{eq:H-ae} and (\ref{eq:Slater-det1}).
By definition, electrons in the {\Inact} space are
constrained to occupy, in the mean-field sense,
the orbitals defined by a lower-level of theory,
such as the canonical orbitals in
HF, DFT, or natural orbitals determined from
an approximate CI calculation.
This implicitly imposes an orthogonality constraint on the \Act-space orbitals.

The AFQMC formalism outlined in the previous section
will need to be modified to implement
this approach.
The {\Inact} orbitals in the AFQMC determinants [\Eq{eq:Slater-det1}]  should be frozen
during the random walks, and orthogonality conditions should be imposed, for numerical stability, on the active electrons.
In addition, the HS operators $\hat{\mathbf{v}}$ need to be modified to
act only in the {\Act} sector of the $N$-electron Hilbert space,
and only the {\Act} orbitals should appear in the force bias $\bm {\bar{\sigma}}[\phi]$ in \Eq{eq:HS-FB}.

An alternative and perhaps more elegant approach is to define a
frozen-orbital Hamiltonian $\Hopfc$ that acts only on the
{\Act} sector of the Hilbert space.
The goal is to have the  {\Inact}
orbitals appear explicitly, if at all, only in one-body operators acting on the {\Act} space.
(They do not appear, for example, in the effective core potential or norm-conserving PP formulations.)
The derivation of $\Hopfc$ proceeds from
a separability approximation of the many-body wave function:
\eql{eq:FCWF}
{
    \Psi \approx \mathcal{A}(\Psi_\Inact \Psi_\Act)
    \,.
}
where the wave functions $\Psi_\Inact$ and $\Psi_\Act$ are assumed to be mutually orthogonal and
individually antisymmetrized and normalized.
The antisymmetrizer $\mathcal{A}$ permutes electrons between $\Psi_\Inact$ and $\Psi_\Act$.
This separation allows the energy of the total system to be effectively mapped onto an equivalent system
involving only the {\Act} electrons:
\eql{eq:E_fs_sep}
{
    E=  \ME{\Psi}{\Hop}{\Psi}
     =    \ME{\Psi_\Act}{\Hopfc}{\Psi_\Act}
    \,,
 }
where
the frozen orbital Hamiltonian $\Hopfc$ is given by \cite{Kahn1976,Huzinaga1991}
\eql{eq:H-fc}
{
    \Hopfc
& = \sum_{ij \in \Act} K_{ij}^{} \Cc{i} \Dc{j}
    + \Half\sum_{ijkl \in \Act}
      V_{ijkl}^{} \Cc{i} \Cc{j} \Dc{k} \Dc{l}
    \\
& + \sum_{ij \in \Act}
    V^{\textrm{\Inact-\Act}}_{ij} \Cc{i} \Dc{j}
+ E^\Inact
    \,.
}
The first term includes the kinetic energy and all one-body external potentials acting on the {\Act} electrons, and the second term
includes the two-body Coulomb interactions among them.
The third term
 is a one-body interaction that represents the interaction between the inactive and active orbitals.
It includes Coulomb and exchange interactions between the {\Act} and {\Inact} electrons;
it is formally identical to a non-local PP.
The fourth term is a constant energy, which represents all interactions among the {\Inact} electrons.
(If HF is used as the low level theory, $E^\Inact$ has the form corresponding to a closed shell determinantal wave function of the {\Inact} electrons. \cite{Huzinaga1991})

As a result of the mapping defined by \Eqs{eq:E_fs_sep} and (\ref{eq:H-fc}),
all of the AFQMC formalism described in Section~\ref{sec:FC-AFQMC-review} can be immediately applied.
Thus, the local energy, with $\Hop \to \Hopfc$ in \Eq{eq:Elocal-def}, is evaluated using a
trial wave function $\PsiT$ and random walkers $\phi$, both of which now depend only on the {\Act} electrons and orbitals.
Similarly, the force bias  $\bm{\bar{\sigma}}$ in \Eq{eq:HS-FB} is evaluated using HS one-body operators $\hat{\mathbf{v}}$ that
are obtained only from the two-body  \Act-space $V_{ijkl}$ matrix elements in \Eq{eq:H-fc}.

In the FC applications to atoms and molecules with GTO basis (\Sec{sec:FC_atoms}),
we use the partitioning provided naturally by the
restricted closed- or open-shell HF
orbitals
of the system being studied.
The HF core orbitals defines the {\Inact} space,
and \Eq{eq:H-fc} is expressed in the basis of
the  valence and virtual orbitals.
(This convention is also widely adopted in correlated quantum chemistry
FC calculations.)
In our calculations we have sometimes used other types of wave function
as $\PsiT$,
such as the DFT, unrestricted HF (UHF), or
the complete active space self-consistent field (CASSCF) wave functions.
For the FC approximation to be valid, the calculation results should be
insensitive to the small variations in the
{\Act-\Inact} partitioning defined by the core orbitals of
these wave functions.
We find that this holds true for our calculations.
Substituting the UHF or CASSCF wave function core orbitals
with those from HF
results in only small changes ($\sim 3$ meV)
in the total energy,
which is negligible for our purposes.

In the downfolding applications (\Sec{sec:FC_downfolding}),
the single-particle basis consists of
the eigenstates from a DFT band structure calculation.
As described later,
a simple truncation scheme is introduced, which will systematically
converge the basis set to the full basis limit
(which is just a unitary transformation from the original
planewave basis).
In the case of spin-polarized DFT calculations, where the spin-up and
down electrons can have different spatial orbitals,
we choose to use the majority-spin orbitals as the basis functions
for both spin sectors.
Additional errors introduced by this choice can thus be considered
as a part of the basis truncation error, which vanishes
in the the limit of full basis.

\section{Benchmarking frozen orbital AFQMC: frozen-core calculations for atoms and molecules}
\label{sec:FC_atoms}

In QMC calculations, a satisfactory treatment of core electrons has not been realized,
despite the absolute necessity of
using some form of a PP as calculations move toward heavier elements and 
ever larger scales. The most commonly used form is atomic pseudopotentials, which are
 usually constructed for reference atomic configurations.
The transferability of the PP across many target systems is challenging to
determine, and systematic accuracy is very difficult to achieve.
In addition, most PPs used in QMC calculations are of the single projector (one per angular momentum channel),
norm-conserving type, which tends to further limit transferability. The frozen-core approach
offers a significant step forward. 
It retains all the advantages of using PPs, namely the reduction of system size by eliminating
core electrons, the change of energy scale and hence the reduction of statistical and time-step errors,
while allowing much better transferability.
The FC Hamiltonian is obtained
 for each target system, using canonical orbitals from a lower level of theory,
 with no additional algorithmic layers.

The
AFQMC frozen-orbital implementation is first tested
by comparisons with standard quantum chemistry methods
for atoms and small molecules, using GTO basis sets.
We compare with
exact full configuration interaction (FCI) calculations, which are feasible for
small systems.
This constitutes a rigorous test of our methodology,
since both AFQMC and FCI use the same
underlying FC Hamiltonian,
\Eq{eq:H-fc}, expressed in the basis of
the orthonormal HF valence and virtual orbitals.
Using larger basis sets, we also calculate ionization potentials
for the transition metal atoms Co and Zn and
compare to experimental values.

\begin{table}[tbp]
\caption{\label{tbl:FCI-benchmarks}
Total energies (hartree atomic units)
for several atoms and small molecules.
FC AFQMC results are compared with exact values from FCI.
The AFQMC statistical error bars are on the last digit and are indicated in parentheses.
The basis sets and the 
the number of FC orbitals ($M_c$)
are indicated.
$\PsiT$ is the trial wave functions used in AFQMC.
}
\begin{tabular}{lccclrr}
\hline\hline
          &  Basis      &  $M_c$ &$\PsiT$&     \quad{}AFQMC        &     FCI         \\
\hline
Be        &  6-31G*     &   $1$  &HF&   \phantom{$00$}$-14.6116(2)$  &   $-14.6134$  \\
Be        &  6-31G*     &   $1$  &CASSCF&   \phantom{$00$}$-14.61361(1)$  &  \\
Li$_2$    &  cc-pVDZ    &   $2$  &HF&   \phantom{$00$}$-14.9017(1)$  &   $-14.9005$  \\
HF        &  cc-pVDZ    &   $1$  &HF&  \phantom{$0$}$-100.2020(1)$  &  $-100.2011$  \\
Zn        &  6-31G
                        &   $9$  &HF& $-1777.6771(2)$  & $-1777.6774$  \\
Zn        &  6-31G
                        &   $9$  &CASSCF& $-1777.6775(5)$  &   \\
Zn$^+$    &  6-31G
                        &   $9$  &HF& $-1777.3705(4)$  & $-1777.3700$  \\
Zn$^+$    &  6-31G
                        &   $9$  &CASSCF& $-1777.3706(5)$  &  \\
\hline\hline
\end{tabular}
\end{table}

\subsection{Total energy comparisons with exact results}

Table~\ref{tbl:FCI-benchmarks} compares AFQMC
total energies with exact results for
some atoms and
molecules.
In Be, HF, and Li$_2$, only the $1s$ states of the Be, F, and Li atoms are frozen,
while in Zn and Zn$^+$,
core states through the $3p$ shell are frozen,
for a total of 9 inactive orbitals.
For most of the systems, AFQMC results were obtained using single-determinant HF $\PsiT$.
Multi-determinant $\PsiT$ derived from
CASSCF
were also used in some cases for comparison.
The Hamiltonian matrix elements and HF wave functions were generated using
a modified {\NWCHEM}\cite{NWChem-6.0} code.
The CASSCF wave functions were generated using
either the {\NWCHEM} or {\GAMESS}\cite{Gamess} quantum chemistry package.
{\GAMESS} is used to calculate the FCI energies.

The AFQMC energies in Table~\ref{tbl:FCI-benchmarks} are in good agreement with the exact results.
Those using a single determinant HF $\PsiT$ have systematic errors typically less than
$\sim 1$~mHa
($0.0272$~eV), well within chemical accuracy.
The largest discrepancy for HF  $\PsiT$  occurs in Be, $\sim 2$ mHa.
This is because of the near-degeneracy of the $2s$ and $2p$ levels in the Be atom.
Using a CASSCF $\PsiT$ of four determinants brings the calculated AFQMC energy
to within only 0.2 mHa of FCI.
For Zn and Zn$^+$, the AFQMC results are insensitive to the $\PsiT$ used.

\begin{table}[thbp]
\caption{\label{tbl:IP-Co-Zn}
Ionization potentials (in eV) of Co and Zn computed using FC AFQMC, compared 
with experimental results. AFQMC results are shown using two different 
trial wave functions: HF and CASSCF. For comparison, 
HF and CCSD(T) results are also shown. 
All calculated results have been extrapolated to the complete basis set limit, as described in the text.
The final AFQMC statistical error bars are on the last digit and are shown in parentheses.
Experimental results have been adjusted to
remove spin-orbit effects (\Ref{Balabanov2005}).
}
\begin{tabular}{lccccccc}
\hline\hline
&                      &           &           &          &  AFQMC/      & AFQMC/       &        \\
&                      &  HF       &  CCSD(T)  &          &  HF          & CASSCF       &  Expt. \\
\hline
\multicolumn{2}{l}{\textbf{Cobalt}}
                       &  $8.30$   &  $7.89$   &          &  $7.73(4)$   &  $7.80(4)$   & $7.87$ \\
\multicolumn{2}{l}{\textbf{Zinc}}
                       &  $7.79$   &  $9.37$   &          &  $9.47(6)$   &  $9.43(4)$   & $9.39$ \\
\hline\hline
\end{tabular}
\end{table}

\subsection{Transition metal atom ionization potentials:
Comparison with experimental results}

We now benchmark the FC AFQMC
in realistic calculations using large basis sets. 
With this formalism, scalar relativistic effects are included straightforwardly in the Hamiltonian
using the Douglas--Kroll--Hess (DKH) approximation.
\cite{Douglas1974,Hess1985,Hess1986}
We compare the ionization potentials (IP) for Co and Zn with
experiment
in Table~\ref{tbl:IP-Co-Zn}.
In Co,  both the atomic and singly ionized ground states have partially-filled
$d$-shell configurations
($3d^7 4s^2$ and $3d^8 4s^0$, respectively);
while both Zn and Zn$^+$ atom have a completely filled $3d^{10}$ shell as well as
spherical $4s^2$ and $4s^1$ configurations, respectively. These atoms
 thus represent
a spectrum of transition metal characters.
Unlike the Zn calculations in Table~\ref{tbl:FCI-benchmarks}, which are small systems to 
compare with exact results, we use a small frozen core
in the calculations here. Only the innermost $1s$, $2s$, and $2p$ core states are
frozen (5 inactive orbitals).
The use of a small core allows the $3s$ and $3p$ electrons to be fully correlated along with $3d$
electrons, which is important for accurate results, since the radial
extent of the $3s$, $3p$, and $3d$ orbitals are similar.
The use of large, relativistic GTO basis sets\cite{Balabanov2005}
is crucial to achieve
systematic extrapolation\cite{Helgaker1997,Purwanto2011} of the calculated results
to the complete basis set (CBS) limit.
\footnote{
We used triple- and quadrupole-zeta
correlation consistent core-valence basis sets (cc-pwCV$x$Z-DK with $x = 3, 4$),
which were designed for use with the DK scalar-relativistic
treatment.
}
The multi-determinant $\PsiT$ used in the AFQMC/CASSCF calculations typically consists of
$\sim 10 - 40$ determinants; 
they are obtained by taking the ones with the largest weights, and they account for $\gtrsim 99\%$ of the total weight of the full CASSCF wave function.
\cite{AlSaidi2007b,Purwanto2008,Purwanto2009_C2}
We have verified the quality of the CASSCF $\PsiT$ by
increasing the size of the active space in the CASSCF calculation
and ensuring that the AFQMC energies with different CASSCF $\PsiT$'s
do not change significantly.

Calculated IP results from AFQMC, extrapolated to the CBS limit,
are shown in Table~\ref{tbl:IP-Co-Zn}.
Also shown are the corresponding results from the coupled-cluster
singles and doubles and perturbative triples
[CCSD(T)] method.
The AFQMC/CASSCF results are seen to be in
excellent agreement with experiment,
as are the CCSD(T) values.
The benchmark results demonstrate the accuracy of FC AFQMC to be comparable
to the best FC quantum chemistry methods, consistent with earlier results from many all-electron 
or ECP calculations.
\cite{AlSaidi2006b,AlSaidi2007b,%
Purwanto2009_C2,Purwanto2011} %
Using the HF trial wave function, the AFQMC/HF values are as good for Zn but slightly worse for Co.
This can be traced to the {\Coplus} mean-field HF solution, which
predicts an incorrect ground state, quintuplet $3d^7\,4s^1$, rather than
tripet $3d^8\,4s^0$.
As seen in the table, the Co ionization potential is overestimated at the HF level.

\section{Downfolding and pseudopotential-free AFQMC calculations in solids}
\label{sec:FC_downfolding}

In this section, we discuss a more general
application of the frozen-orbital method in extended systems,
where the broader aim is to greatly reduce
the relevant degrees of freedom and/or
the number of explicitly correlated electrons.
The idea is
related to the downfolding method, which focuses on the
low-energy, physically relevant, sector of the Hilbert space, while downfolding  high-energy states to obtain an effective Hamiltonian that acts only in the low-energy subspace.
We present an
example, using the frozen orbital approximation, to
replace high energy planewave basis states in favor of a more compact representation of lower energy
basis functions. Since the downfolding and generation of one- and two-body matrix elements are
done at the mean-field level, the approach also affords the opportunity to eliminate most of the errors associated with the pseudopotential in the subsequent many-body calculations, as we describe
below.

Planewave basis sets
are commonly used to describe
extended systems.
Calculations using a planewave basis are appealing
because the basis is complete
and convergence to the CBS limit is straightforward,
using only a single cutoff energy parameter $\Ecut$.
Planewave basis sets can be inefficient, however. Since  the basis is
unbiased and does not build in information about the specific system being studied,
large basis sets may be required, especially in supercell calculations. Basis sets constructed from
localized orbitals, on the other hand, can be tailored to the physics of a particular system, although
convergence to the CBS limit is not as straightforward. There are many possible choices for
a local orbital basis, including Wannier functions, GTO, Slater-type orbitals (STO), or numerical basis sets.
For the present application, we choose a simple transformation
from the planewave basis to the basis of Kohn-Sham (KS) orbitals,
where the associated KS band energies (denoted as $\epsilon$) are
used to truncate the new basis according to a specified
energy cutoff.
Other choices are likely to be more efficient for studies of extended systems, but this simple
approach illustrates the concept.
Moreover, it is systematically improvable by simply increasing the band cutoff; in the limit
of including all bands, it is just a unitary transformation from the full planewave basis.
For certain applications, involving similar atomic arrangements, where favorable cancellation
of errors may be expected,
this straightforward approach can lead to significant savings.  

\begin{figure}[htbp]
\includegraphics[scale=0.70]{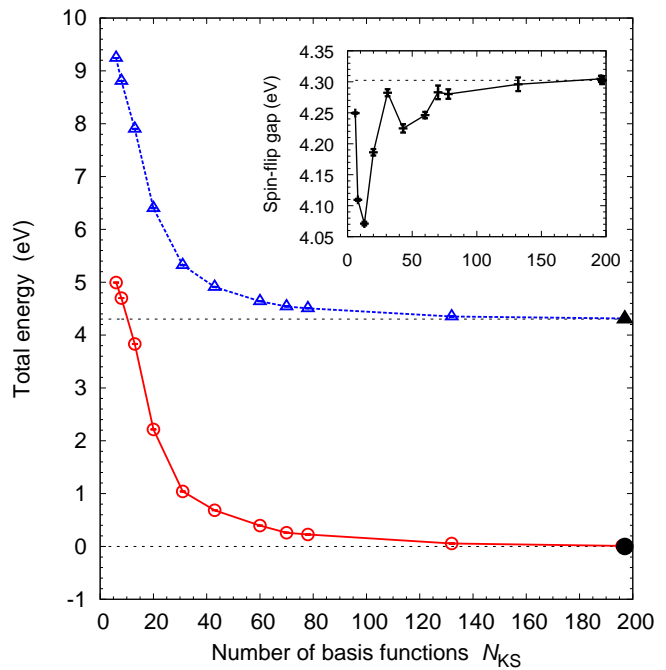}
\caption{\label{fig:Si-gap}
(Color online)
Convergence of the AFQMC total energies and spin gap 
as functions of the number of KS bands ($\Nks$).
Total energies of the ground (excited/spin-flip) state in the silicon primitive cell are shown as
open circles (triangles), respectively.
The filled symbols denote the planewave AFQMC results.
Total energies are shifted such that the planewave AFQMC ground-state energy
is zero.
Shown in the inset is the corresponding spin gap. QMC statistical error bars are not visible,
as they are much smaller than the symbol size.
}
\end{figure}

We first illustrate the convergence behavior of the KS basis by
calculating the spin gap in bulk silicon.
A primitive cell is considered at a single $\kvec$-point (Baldereschi \cite{Baldereschi1973}).
Using each set of truncated KS orbitals as the basis set,
we construct the Hamiltonian in \Eq{eq:H-ae} and calculate the AFQMC energy.
We increase the number of the KS orbitals ($\Nks$) until it reaches
the full number of planewaves.
These calculations are then operationally equivalent to AFQMC with a GTO basis \cite{AlSaidi2006b};
all information about
the structure and periodicity of the cell  is reflected only in the values of the matrix elements.
The results are shown  in
\Fig{fig:Si-gap} and compared with the full planewave results which are obtained
with planewave-AFQMC calculations \cite{Suewattana2007,Purwanto2009_Si}.
The DFT and planewave AFQMC calculations used a norm-conserving PP
with a planewave cutoff of $12.25$ Ry;
the KS-basis AFQMC does not employ the FC approximation
in this test, treating all the valence electrons as defined by DFT.
The total energies and the spin gap converge to the full planewave
limit, as expected.
The total energies shows a slow, monotonic convergence behavior as a
function of $\Nks$.
However, the spin-flip gap computed using the KS basis is much closer to the
planewave gap even with a small number of KS bands.
For example, at $\Nks \sim 40$, the total energy is almost $1$\,eV away from
the basis set limit, but the gap is only $\sim 0.07$\,eV away.
The KS gap, however, does not converge monotonically as a function of
$\Nks$.

We next apply the 
downfolding
technique to a more correlated extended system.
We will focus on crystalline MnO, and will demonstrate
that the method provides a treatment of a strongly correlated system using a realistic 
model Hamiltonian and free of pseudopotential errors.
As a simple prototype for Mott insulators,
MnO poses a major challenge for theoretical methods,
with the presence of localized $3d$ electrons.
At ambient temperatures and pressures,
MnO chemically orders in a rocksalt
crystal structure.
At temperatures lower than 118 K,\cite{Yoo2005}
an antiferromagnetic ordering (type-II AFM) sets in
accompanying rhombohedral distortion of the crystal.
High-pressure experiments\cite{Kondo2000,Yoo2005}
have presented evidence of a simultaneous
structural and magnetic phase transitions.
At $\sim 105$ GPa, the crystal volume collapses by $\sim 20\%$,
accompanied by a local magnetic moment collapse.
Various DFT calculations yield varying predictions for the nature of the
transition, \cite{Kasinathan2006}
with different predictions of the
transition pressure and whether the transition is from metal to insulator.

We use the primitive rocksalt MnO unit cell
with periodic boundary conditions as the model system to
compute the high-spin to low-spin state vertical energy gap.
As the primitive cell contains an odd number of electrons, we represent the low-spin
state
by the $S=1/2$ state and the 
ferromagnetic phase
by the $S=5/2$ state. 
The
spin gap is computed as
\eql{eq:NM-FM-gap-def}
{
    \Delta E \equiv E_{S=1/2}^{} - E_{S=5/2}^{}
    \,.
}
All calculations were done near the experimental volume, corresponding to
a rocksalt cubic lattice constant of $a = 8.4\,\textrm{\AA}$.
Both the DFT and QMC calculations were done at the $L$ point in the
Brillouin zone.
We have chosen a small simulation cell here so that high resolution calculations can be 
easily done.
Our goal is thus not to compare
with experiment or other theoretical results,
but rather to study the accuracy of the downfolded many-body Hamiltonian as a function
of band cutoff,
comparing to AFQMC results using the full planewave basis.

A secondary goal of this work is to show how the frozen-core approach combined with downfolding
can largely remove the error from the use of conventional norm-conserving PPs.
As discussed earlier, this error poses a significant problem in QMC calculations.
In systems containing $3d$ transition metal atoms, for example,
relatively hard pseudopotentials are typically constructed such that the semicore  $3s$ and $3p$ states are included as
valence states (Ne-core PP), which is especially important for the early $3d$ atoms.
Nevertheless, Ne-core PPs have been found to
introduce unsatisfactorily large errors.
\cite{Kolorenc2007}
We show below that this problem can be largely ameliorated with our approach. 
A He-core PP (retaining the $2s$ and $2p$ as valence states) is used for the transition metal atom to generate 
the KS basis set.
In the 
subsequent AFQMC calculations with the downfolded Hamiltonian, the $2s$ and $2p$
states are treated with the 
FC approximation, freezing them at the DFT level using the orbitals
derived from the solid rather than from an atomic calculation.

The PP DFT calculations were done with the {\ABINIT}\cite{Abinit} package,
using the generalized gradient Perdew--Becke--Ernzerhof exchange-correlation functional;
{\PWSCF}\cite{QEspresso} was used for the
projector augmented wave (PAW) calculations;
all-electron linearized augmented planewave (LAPW) calculations were done with {\ELK}\cite{Elk}.
All our norm-conserving PPs assume the Kleinman-Bylander (KB) form,
and were generated
using the {\OPIUM} package.\cite{Opium}
A standard norm-conserving He-core PP is used for the O atoms.%
\footnote{
    The O PP was generated with the following parameters:
    $r_c = 1.1$ {a.u.} for the $s$ and $p$ channels,
    with the $s$ channel being the local projector.
}
Both Ne- and He-core Mn PPs used a single-projector for each angular momentum channel
(as is almost universally done for norm-conserving PPs) and were designed for planewave
cutoffs of $100$ and $1600$ Ry, respectively.%
\footnote{
    The Ne-core Mn PP has $r_c$'s of
    $1.15$, $1.15$, $1.5$ {a.u.} for the $s$, $p$, and $d$ channels
    (using $p$ channel as the local projector).
    The He-core Mn PP has $r_c$'s of
    0.36 0.305 0.5
    $0.36$, $0.305$, $0.5$ {a.u.} for the $s$, $p$, and $d$ channels
    (using $s$ channel as the local projector).
    Scalar relativity is included in the construction of both PPs.
}
Direct use of He-core PPs in QMC is costly because of the
hardness of the PP (exceedingly high cutoff) and the inclusion of the eight additional $2s$ and $2p$ electrons
per atom, and is almost never done
(although multiple-projector norm-conserving PPs, or even the more accurate PAW, could be
implemented in AFQMC).
The same number of KS orbitals were used for both the low- and high-spin states.

\begin{figure}[thbp]
\includegraphics[scale=0.70]{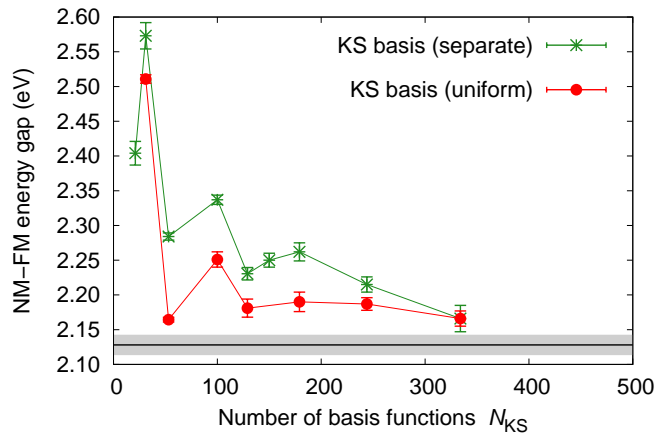}
\caption{\label{fig:MnO-1x1x1-basis-convg}
(Color online)
Convergence of the AFQMC high--low spin gap
in MnO
as a function of $\Nks$.
The KS gaps are shown as points with the statistical error bars; lines are only
to aid the eye.
The two curves correspond to two different approaches in
defining the truncated KS basis, as described
in the text.
The energy gap from the full planewave AFQMC calculation
is shown as a solid black horizontal
line with the gray shading representing its statistical error.
}
\end{figure}
Figure~\ref{fig:MnO-1x1x1-basis-convg} shows the convergence of the spin-gap $\Delta E$
with respect to $\Nks$.
These calculations were done with the Ne-core PP.
The spin gap value from the full planewave basis AFQMC
is also shown.
The graph shows the spin gaps computed in two slightly different downfolding approaches.
In the first approach
(``separate KS basis''),
the KS basis used in AFQMC is constructed from
the corresponding DFT calculation for each magnetic state.
This results in different downfolded Hilbert spaces for the two calculations,
which have different convergence rates to the full planewave basis limit.
In the second approach
(``uniform KS basis''),
only one KS basis is used in AFQMC calculations for all the magnetic states;
in this case, we expand all the Hamiltonians and wave functions
in terms of the majority spin KS orbitals of the $S=5/2$ magnetic state.
As seen in Fig.~\ref{fig:MnO-1x1x1-basis-convg}, the second approach
converges more rapidly due to better cancellation of basis-set errors.
The difference of the two approach diminishes systematically as $\Nks$ approaches
the full planewave basis limit, as expected.
Since it is highly desirable to use a downfolded basis that will
converge rapidly to the full planewave basis limit,
we will
use the second approach
in the subsequent calculations.

The results in \Fig{fig:MnO-1x1x1-basis-convg}
show that the calculated spin gap
is already in good agreement
with the full planewave result
at $53$ basis functions.
This represents only a small fraction of the full planewave basis (2488 basis functions).
Similar to the result in silicon,
the convergence of the spin gap to the full basis set limit is not monotonic,
showing the largest deviation ($\sim 0.12$~eV) at $\Nks = 100$,
although the error in the gap is well below $0.1$~eV beyond $100$ basis functions.
These results indicate that it is possible to construct realistic models from the 
downfolding approach which are much simpler than the 
full Hamiltonian and capable of giving quantitatively accurate results.

\begin{figure}[tb]
\includegraphics[scale=0.30]{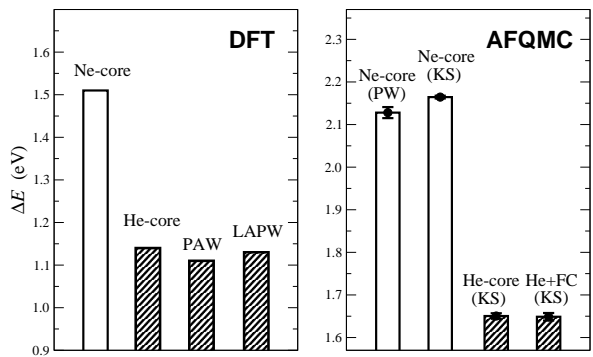}
\caption{\label{fig:E-convg-big}
Calculations of the energy gap $\Delta E$ between the high- and low-spin phases in MnO.
DFT results are shown on the left panel,
while AFQMC results on the right.
Downfolded Hamiltonians lead to accurate results in AFQMC, as seen from the
comparison with the planewave calculations.
The single-projector, Ne-core PP
is inadequate, as illustrated at both the DFT and the QMC levels of theory.
AFQMC with FC Hamiltonian (He+FC) is in excellent agreement 
with the ``all-electron'' result.
Actual numerical values are presented in Table~\ref{tbl:E-convg-big}.
}
\end{figure}
We next demonstrate the significant error of the standard single-projector, norm-conserving Ne-core
PP as employed in many-body simulations, and how it can be removed in the current approach with little additional computational 
cost.
The PP effects on $\Delta E$ are shown in \Fig{fig:E-convg-big}.
The left panel compares Ne-core and He-core PP DFT calculations with
PAW and with all-electron LAPW values.
The results are tabulated in
Table~\ref{tbl:E-convg-big}.
\begin{table*}[!bhtp]
\caption{\label{tbl:E-convg-big}
Accurate determination of the energy gap $\Delta E$ (in eV) between the high-
and low-spin phases
with downfolded Hamiltonians
and FC calculations, and the inadequacy of single-projector pseudopotentials.
AFQMC results using Ne-core PP,
He-core PP, He-core PP with FC approximation are compared. Corresponding DFT
calculations are also shown to illustrate consistency.
$N_e$ is the number of electrons in the simulation cell, and
$M$ is the number of single-particle basis functions.
}
\begin{tabular}{lccrccc}
\hline
\hline
                    &          &\multicolumn{2}{c}{Basis set} & Cutoff       &  DFT/GGA     &  AFQMC       \\
        \cline{3-4}
Mn PP               &  $N_e$   & Type     &     $M$   & parameter (Ry)       &  $\Delta E$  &  $\Delta E$  \\
\hline
\hline
Ne-core             &   $21$   & PW       &   $2488$  &  $\Ecut=100$         &    $1.51$    &  $2.13(1)$\phantom{$0$}   \\
                    &          & KS       &     $53$  &  $\epsKS=9.6$        &              &  $2.164(3)$   \\
\hline
He-core             &   $29$   & PW       & $160046$  &  $\Ecut=1600$        &    $1.14$    &              \\
                    &          & KS       &     $57$  &  $\epsKS=9.6$        &              &  $1.651(7)$   \\
\hline
He-core + FC        &   $21$   & KS       &     $53$  &  $\epsKS=9.6$        &              &  $1.649(9)$   \\
\hline
AE                  &   $21$   & PAW\footnote[1]{Ne core was frozen at atomic level}
                                          &   $4584$  &  $\Ecut=150$         &     $1.11$   &              \\
AE                  &   $29$   & LAPW     &    $266$  &  $\Ecut=22$          &     $1.13$   &              \\
\hline
\hline
\end{tabular}
\end{table*}
The DFT calculations show that the Ne-core PP
leads to an overestimation of the gap by almost $0.4$ eV.
By contrast, the He-core PP gap
is in excellent agreement with the all-electron LAPW value.
While the PAW calculation also uses a Ne-core PP,
it uses two (or more) projectors per angular momentum channel
and is seen to reproduce the LAPW result well.
We therefore attribute the poor performance of the
Ne-core norm-conserving PP not to
the underlying frozen core approximation for the Mn $2s$ and $2p$
states, but rather to the deficiency of its being only a single-projector PP.
The excellent transferability of the He-core PP indicates that the single-projector representation is sufficient for
this much harder PP.

The right panel in \Fig{fig:E-convg-big} shows AFQMC results, which are consistent with 
the DFT trends and illustrate clearly the different aspects of the transferability 
issues in a many-body context.
For the Ne-core PP, results are shown from both
the full planewave basis and the KS basis ($\Nks=53$, as described earlier).
For the He-core PP, AFQMC results are shown for both an ``all-electron'' (He-core) Hamiltonian
fully correlating the $2s$ and $2p$ Mn states ($\Nks=57$),
and from a FC Hamiltonian ($\Nks=53$),
where these states are frozen at the DFT level. 
The discrepancy in $\Delta E$ between 
the Ne-core PP and the He-core results increases to
 $\sim 0.5$ eV in the many-body results.
The excellent agreement between the two He-core results indicates the accuracy of the FC approximation
in the many-body calculations.
Accurate and efficient AFQMC calculations are thus achieved at
a cost comparable to a Ne-core PP using the downfolded Hamiltonian
and the FC approximation.

\section{Summary}
\label{sec:Summary}

The frozen orbital and downfolding approach described in this paper
can provide significant computational savings compared
to fully correlating all the electrons in both
molecular and extended systems.
The key idea of identifying a physically important ``active'' subspace
of the full Hilbert space
is already inherent in the standard FC approximation
used in explicitly correlated wave function based many-body quantum chemistry methods.
We have shown how the FC approximation can be implemented in AFQMC.
With a GTO basis set, FC AFQMC treats \emph{exactly the same}
Hamiltonian as standard quantum chemistry methods.
This effectively eliminates the chemically inactive core degrees of freedom
(and electrons) from the calculation,
resulting in greatly increased computational efficiency,
particularly for heavy atoms.
Scalar relativity for such systems can be easily treated
using the DKH approximation to the Hamiltonian.
More generally, 
the downfolding of high-energy basis states
to a physically relevant low-energy sector
can greatly increase the efficiency of this approach in solid-state applications. 
As an example, we have shown in this paper how to effectively eliminate
the error due to single-projector, norm-conserving PPs.
The fact that AFQMC is an orbital-based method
is the key feature that enables
the partitioning and downfolding schemes described here.

The results presented in this paper are a proof of concept
that the downfolding and FC approaches could greatly extend the reach of
many-body calculations to larger and more complex systems.
Clearly, AFQMC applications with these approaches require further study.
For example, the simple downfolding application we have described is based on a single energy cutoff.
The efficiency of the method could be greatly improved by employing
additional physically-based criteria.

The FC approach can also be generalized to other partitioning schemes
of the Hilbert space into active and inactive regions.
In molecular and condensed matter physics, for example,
the active region may sometimes be identified spatially,
corresponding to a localized region where strong electron correlation effects
affect a relatively small number of atoms, while the bulk of the system can be treated
with a lower level of theory.
Various theories of partitioning (see \Ref{Huzinaga1991}, for example) or embedding
\cite{Sven1996,Govind1999,Huang2006,Huang2008,Elliott2010}
have been develop to exploit this locality
and
improve computational efficiency.
\cite{Knizia2013,Beran2012}
AFQMC can also benefit from efficiencies derived from the use of
localized orbital transformations
obtained from Boys\cite{Boys1960} or
Wannier\cite{Wannier1937,Marzari1997} localization.

\begin{acknowledgments}

This work was supported by
DOE (DE-FG02-09ER16046),
NSF (DMR-1006217),
and
ONR (N000140811235; N000141211042).
An award of computer time was provided by
the Innovative and Novel Computational Impact on Theory and Experiment
(INCITE) program,
using resources of the Oak Ridge Leadership Computing Facility (Jaguar/Titan)
at the Oak Ridge National Laboratory,
which is supported by the Office of Science of the U.S. Department of Energy
under Contract No. DE-AC05-00OR22725.
This research is part of the Blue Waters sustained-petascale computing
project, which is supported by the National Science Foundation (award
number OCI 07-25070) and the state of Illinois. Blue Waters is a joint
effort of the University of Illinois at Urbana-Champaign and its National
Center for Supercomputing Applications.
We also acknowledge the computing support from the Center for Piezoelectrics by Design.
The authors would like to thank Eric J. Walter for
many useful discussions, his help in generating pseudopotentials and
performing the PAW and LAPW calculations.

\end{acknowledgments}

\bibliography{AFQMC-bib-entries}

\end{document}